\documentclass[smallextended]{svjour3} 
\usepackage[nocompress]{cite}
\usepackage{natbib}
\usepackage{cite}
\input{setup}
\usepackage[colorlinks,bookmarksopen,bookmarksnumbered,citecolor=red,urlcolor=red]{hyperref}
\usepackage{ragged2e}
\usepackage{multirow}
\usepackage{listings}
\usepackage{enumitem}
\usepackage{xcolor}

\definecolor{ao}{rgb}{0.0, 0.0, 0.8}
\definecolor{cyan4}{RGB}{0,139,139}
\definecolor{navy}{RGB}{0,0,128}
\usepackage[toc,page]{appendix}

\definecolor{custom-gray}{cmyk}{0, 0, 0, 0.7, 1.00}
\usepackage[most]{tcolorbox}
\definecolor{custom-gray}{cmyk}{0, 0, 0, 0.7, 1.00}
\newtcbtheorem[no counter]{Summary}{\hskip-0.97em}{enhanced,drop shadow={black!50!white},
 coltitle=white,
 top=0.15in,
 attach boxed title to top left=
 {xshift=1.5em,yshift=-\tcboxedtitleheight/2},
 boxed title style={size=small,colback=custom-gray}
}{summary}
\begin{document}

\title {Toward an Evidence-Based Template for Agent Instruction Files: An Empirical Study of Adoption, Consistency, and Maintenance}
\author{Mehedi Hasan Shanto \and Md. Anaytul Islam$^{\delta}$ \and Muhammad Asaduzzaman \and Alioune Ngom}
\titlerunning{ Toward an Evidence-Based Template for Agent Instruction Files}


\institute
{Mehedi Hasan Shanto \and Muhammad Asaduzzaman \and Alioune Ngom  \at
              School of Computer Science \\
              University of Windsor, Ontario, Canada\\
              \email 
              {\{shanto1, masaduzz, angom \}@uwindsor.ca}
\and Md. Anaytul Islam \at
             Independent Researcher\\
              \email {mislam44@lakeheadu.ca}
}
\date{Received: date / Accepted: date}

\maketitle
\begin{abstract}

AI coding agents rely on repository-level instruction files to understand project conventions, development workflows, and constraints. However, different agents support different instruction formats, and limited empirical evidence exists about how these files are adopted, what information they contain, whether coexisting formats provide consistent guidance, and how they are maintained. We conduct an empirical study of 3,387 instruction files from 1,000 GitHub repositories across four formats: AGENTS.md, CLAUDE.md, GitHub Copilot instructions, and Cursor rules. We find that 42.2\% of instruction-bearing repositories in the higher-popularity cohort maintain multiple formats. Among 386 manually examined pairs of corresponding sections, 16.1\% provide divergent guidance. Our maintenance analysis further shows that 40.4\% of the analyzed files have only one recorded modification event and that maintenance patterns differ across formats. Based on category prevalence and category addition ratios (CAR), we derive an evidence-based template containing ten Required, four Gap, and two Recommended categories. In an exploratory generation experiment, explicit template guidance increases structural category coverage from 18.7\% to 82.8\%. We further introduce \textbf{AGIT-Instruct}, a benchmark for evaluating instruction-file generation through structural coverage and content alignment. Across four generation configurations, retrieval-augmented generation achieves the highest full-file ROUGE-1 F1 of 0.459. A manual assessment of 278 generated category instances finds that 19.4\% fully preserve and 61.2\% partially preserve developer-written intent. Our findings provide empirical guidance for organizing and maintaining agent instruction files and demonstrate that structural completeness alone is insufficient for evaluating automatically generated instructions. Our replication package is available at \url{https://github.com/mehedisresearch/AGIT-Template}.
\end{abstract}

\maketitle


\section{Introduction}
\label{sec:introduction}
Large language models (LLMs) are increasingly integrated into software development through AI coding agents that can inspect repositories, modify source code, execute tests, and create pull requests with limited human intervention~\citep{Li2026AIDev,Hassan2025AgenticSE}. These agents perform multi-step development
tasks that require knowledge of
repository-specific practices, including
build procedures, testing workflows, coding
conventions, architectural constraints,
and restrictions on code changes. Much of this information is distributed across source code, documentation, configuration files, and development workflows. Providing agents with the appropriate repository context has therefore become an important concern in agentic software engineering ~\citep{Mei2025ContextEngineering}.

One mechanism for providing this context is the \emph{agent instruction files}. Such files contain repository-specific guidance that coding agents can consult when performing development tasks. These files commonly describe build and test commands, coding conventions, repository structure, and restrictions on code modifications~\citep{Chatlatanagulchai2025Manifests, Chatlatanagulchai2025AgentREADMEs}. Different coding tools have introduced different instruction-file formats, including \texttt{AGENTS.md}, \texttt{CLAUDE.md}, GitHub Copilot instructions, and Cursor rules. Although these formats serve similar purposes, they differ in their conventions and configuration mechanisms ~\citep{ConfiguringAgenticTools2026,Jiang2026Cursor}. Consequently, repositories that use multiple coding agents may need to maintain several instruction files simultaneously.

The coexistence of multiple instruction file formats introduces a potential maintenance challenge. Galster et al.~\citep{ConfiguringAgenticTools2026} found that developers configure coding agents through different repository-level artifacts, with distinct configuration practices emerging across tools. Different files may legitimately contain agent-specific guidance, and their presence alone does not imply duplication or inconsistency. However, when multiple files describe the same repository practice, their instructions must remain aligned as the project evolves. For example, a repository may maintain both an AGENTS.md file containing general development conventions and a CLAUDE.md file providing guidance for Claude Code. If both files describe the repository's testing practices, differences in their instructions could leave agents with different information about the same development task. If one representation changes while another does not, different agents may receive different guidance for the same development activity. Conversely, apparently different sections may provide complementary information rather than conflicting instructions. Such differences may reflect tool-specific requirements rather than errors. Understanding multi-format repositories therefore requires examining not only whether instruction files coexist, but also what information they contain, how their guidance relates across formats, and how these artifacts are maintained over time.

Recent empirical studies have begun to characterize agent instruction files and related configuration artifacts. Chatlatanagulchai et al.~\citep{Chatlatanagulchai2025Manifests} investigated Claude Code manifests and characterized the information developers provide to coding agents. Their subsequent work examined agent instruction files more broadly ~\citep{Chatlatanagulchai2025AgentREADMEs}. Galster et al.~\citep{ConfiguringAgenticTools2026} studied configuration practices for agentic AI, while Jiang and Nam~\citep{Jiang2026Cursor} investigated Cursor rules. Other studies have examined instruction file quality and maintenance. Ardic et al.~\citep{Ardic2026SCAM} investigated issues in \texttt{AGENTS.md} files, and Voria et al.~\citep{Voria2026Evolution} studied how developers maintain and evolve agent instructions. These studies establish an important foundation for understanding agent instruction practices. 

Despite this progress, several questions remain. First, prior studies often focus on individual formats, configuration mechanisms, or aspects of instruction quality, leaving limited evidence about how different instruction formats coexist within the same repository. Second, the presence of similar topics does not establish that corresponding sections provide equivalent guidance. Two files may both contain a testing section while prescribing different commands or constraints. Third, existing work has provided limited comparative evidence about how instruction files are maintained across formats and which categories are predominantly added or removed during their evolution. Finally, content prevalence and maintenance behavior have rarely been considered together when determining which instruction categories should be included in a reusable template. These gaps also affect automated instruction generation. Prior research has investigated the role of repository context and instructions in agent behavior~\citep{Mei2025ContextEngineering, Lulla2026Impact,Gloaguen2026Evaluating}. However, generating an instruction file requires more than producing plausible text. The generated file should cover relevant instruction categories and communicate guidance that is consistent with developers' intentions. A model may produce fluent instructions while omitting important categories. Conversely, a generated instruction file may cover many expected categories while providing generic, incomplete, or inaccurate content. Therefore, evaluating instruction generation requires considering both \emph{structural coverage} and \emph{content alignment}. In particular, the presence of a category does not establish that its instructions preserve the complete developer intent.

To investigate the above-mentioned issues, we conduct an empirical study of 3,387 agent instruction files from 1,000 GitHub repositories selected from the AIDev dataset~\citep{Li2026AIDev}. We examine four instruction formats: \texttt{AGENTS.md}, \texttt{CLAUDE.md}, GitHub Copilot instructions, and Cursor rules. We formulate three research questions (RQs) to understand their adoption, content, and maintenance practices.

\begin{description}
\item \textbf{RQ1: How are agent instruction files adopted across repositories and formats?}
Understanding adoption is important to determine how widely developers use agent instruction files and whether multiple formats coexist within repositories. We find that instruction files are more common in high-popularity repositories (66.2\%) than in low-popularity repositories (50.2\%). Among high-popularity repositories containing instruction files, \textbf{42.2\% maintain multiple formats}, indicating that developers often provide guidance for more than one coding agent.
\item \textbf{RQ2: What information do agent instruction files contain, and how consistent are they across formats?}
Examining their content and consistency is important to identify what guidance developers provide and whether different agents receive compatible instructions within the same repository. We manually code instruction files using the taxonomy of Chatlatanagulchai et al.~\citep{Chatlatanagulchai2025AgentREADMEs} and find that files within the same repository often cover similar topics. However, \textbf{16.1\% of 386 manually examined section pairs provide divergent guidance}. We also identify underrepresented categories, including security, ownership, maintenance, and performance requirements.
\item \textbf{RQ3: How do developers maintain agent instruction files over time?}
Investigating maintenance is important to understand how developers update instruction files as repositories evolve. We analyze modification frequency, ownership concentration, churn ratio, lifecycle patterns, and update triggers. We find that \textbf{40.4\% of the analyzed files have only one recorded modification event}. Maintenance patterns also differ across formats, particularly in whether instruction updates occur independently or alongside other repository changes.
\end{description}

Based on the results of these research questions, we derive an evidence-based template containing 16 categories organized into ten Required, four Gap, and two Recommended categories. The tier assignment combines category prevalence with the category addition ratio (CAR), which captures whether observed category-change events are dominated by additions or deletions. The template is intended as an empirically motivated authoring aid rather than a universally optimal instruction structure: its tiers reflect patterns in the analyzed corpus and do not imply that every category is necessary for every repository. 
We then examine how the template can support automated instruction generation. In an exploratory comparison, explicit template guidance increases structural category coverage from 18.7\% to 82.8\%. To evaluate generation more systematically, we introduce \textbf{AGIT-Instruct}, a benchmark constructed with 1,815 eligible instruction files across 522 repositories. Its repository-disjoint evaluation set contains 62 files and 556 reference category instances. We evaluate GPT-4o-mini using zero-shot prompting and retrieval-augmented generation (RAG), together with CodeLlama-13B in base and QLoRA-adapted configurations. GPT-4o-mini zero-shot achieves the highest structural coverage (82.8\%), while RAG achieves the highest full-file ROUGE-1 F1 (0.459). QLoRA improves CodeLlama-13B's full-file embedding similarity from 0.373 to 0.608 despite substantially lower structural coverage. These results demonstrate that category presence and content alignment capture different properties of generated instructions. 
We further manually examine 278 RAG-generated category instances from 16 instruction categories. Of these, 19.4\% fully preserve the developer-written reference intent, 61.2\% preserve it only partially, and 19.4\% are missing. The predominance of partial matches indicates that generated instructions can capture the general purpose of a section while omitting individual requirements, conditions, or constraints. 

Based on our findings, we provide actionable recommendations for developers, tool providers, and researchers. Developers can use our proposed template to document essential repository guidance and coordinate instructions across multiple formats. Tool providers can develop automated mechanisms to identify overlapping guidance, detect potential inconsistencies, and support timely updates. Researchers can further improve instruction generation by focusing on preserving developer intent rather than merely achieving structural coverage or semantic similarity.

\noindent\textbf{Paper Organization.} The remainder of this paper is organized as follows. Section~\ref{sec:motivating} presents a motivating example, and Section~\ref{sec:design} describes the dataset and research methodology. Section~\ref{sec:results} reports the empirical findings. Section~\ref{sec:template} introduces the evidence-based template and maintenance protocol. Section~\ref{sec:generation} presents AGIT-Instruct and the generation evaluation. Section~\ref{sec:discussion} discusses the findings and their implications, followed by related work in Section~\ref{sec:related_work}, threats to validity in Section~\ref{sec:threats}, and conclusions in Section~\ref{sec:conclusion}.

\section{Motivating Example}
\label{sec:motivating}

We found a motivating example in \textbf{Dawarich}, a popular open-source
location-history tracking application built
primarily with Ruby on Rails. As of October
2026, the repository had approximately
10,600 GitHub stars and 390 forks, making
it a useful real-world example of an
actively maintained project with multiple
agent instruction files.\footnote{
Repository: \url{https://github.com/Freika/dawarich}.
Repository statistics and instruction files
accessed in October 2026.}

Dawarich maintains both \texttt{AGENTS.md}
and \texttt{CLAUDE.md} at the repository
root. Although both files provide
development guidance, they differ in
their formatting instructions.
Specifically, \texttt{AGENTS.md}
instructs agents to check JavaScript
formatting using Prettier
(\texttt{npx prettier --check app/javascript}),
whereas \texttt{CLAUDE.md} requires
Biome for JavaScript and CSS formatting
(\texttt{npx @biomejs/biome check --write}).
Both files also describe testing
procedures involving RSpec and Playwright,
but organize the commands differently.
\footnote{
\url{https://github.com/Freika/dawarich/blob/master/AGENTS.md}
and
\url{https://github.com/Freika/dawarich/blob/master/CLAUDE.md}.}

This example illustrates that multiple
instruction files can cover the same
development activities while prescribing
different workflows. Although such
differences may be intentional, they
can create confusion for coding agents
and require coordinated maintenance
when development practices change.
This motivates our investigation of
instruction file adoption, cross-format
content consistency, and maintenance
practices.



\section{Methodology} \label{sec:design}
\begin{figure*}[ht]
  \centering
  \includegraphics[width=\linewidth]{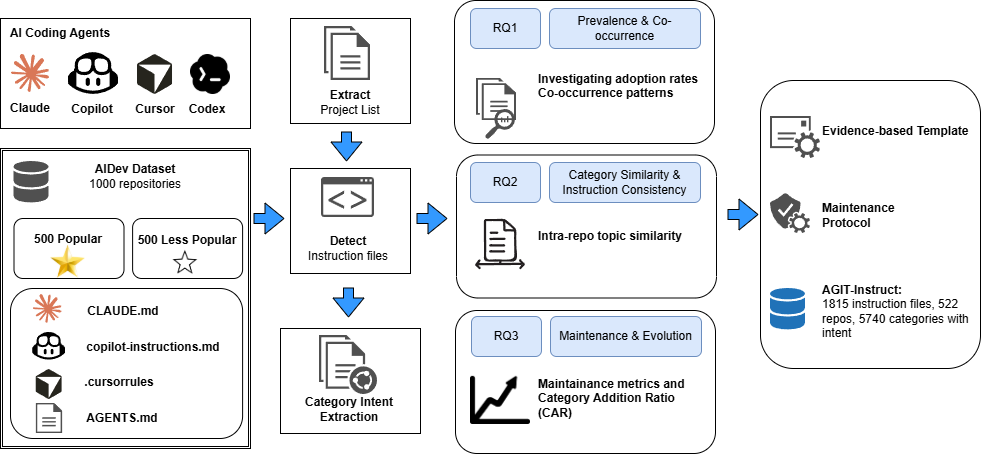}
  \caption{Overview of our study methodology.}
  \label{fig:framework}
\end{figure*}

To answer our three research questions, we collect and process agent instruction files from open-source repositories using the AIDev dataset~\citep{Li2026AIDev}. Figure~\ref{fig:framework} illustrates the overall study pipeline.
We identify repositories that use AI coding agents, detect their
instruction files across four formats, and apply three complementary
analyses covering adoption, content similarity, and maintenance
behavior.

\paragraph{\textbf{Repository Selection.}}
We use the AIDev dataset~\citep{Li2026AIDev}, which contains 932,791 pull requests submitted by different AI coding agents (e.g., OpenAI Codex, GitHub Copilot, Cursor, and Claude Code) across 116,211 open-source GitHub repositories. We divide the repositories into two cohorts based on the median GitHub star count, a commonly used proxy for repository popularity and community engagement~\citep{Borges2016Stars}. The \textbf{top cohort} contains repositories with star counts above the median, while the \textbf{bottom cohort} contains those below the median. We randomly sample 500 repositories from each cohort, yielding 1,000 repositories in total. For each repository, we collect its name, primary programming language, star count, fork count, and license. This stratified sampling strategy enables us to compare agent instruction practices across repositories with different popularity levels.\\
\paragraph{\textbf{Instruction File Detection.}}
Since AIDev \cite{Li2026AIDev} provides pull request metadata rather than the complete content of agent instruction files, we additionally clone the 1,000 selected repositories and extract instruction files directly from their checked-out contents for subsequent analysis. We identify four formats using case-insensitive filename matching: \texttt{AGENTS.md}, \texttt{CLAUDE.md}, \texttt{.github/copilot instructions.md}, and \texttt{.cursorrules}. Each detected file is recorded as a separate observation linked to its repository, relative path, and source format. When a repository contains multiple instruction files, we retain each file separately. We record two repository-level indicators: whether a repository contains at least one instruction file and whether it contains multiple instruction formats. These indicators are computed from unique repository counts, while file-level analyses retain all detected files. Instruction-file adoption therefore describes repository-level presence at the inspected revision. We separately use pull request histories and patches from~\cite{Li2026AIDev} to analyze recorded instruction-file modifications in RQ3.\\
\paragraph{\textbf{Content Category Mapping.}}
\label{sec:taxonomy}
To compare content across the four formats, we map each section of every file to one of 16 semantic categories using the taxonomy of Chatlatanagulchai et al.~\cite{Chatlatanagulchai2025AgentREADMEs}. The 16 categories cover the main topics found in agent instruction files: Do/Don't rules, Build/run/install, Repository structure, Test/CI commands, Code style, Architecture, Domain models, API/contracts, Examples, Tooling agent config, Git/PR conventions, Security/secrets, Performance, Maintenance/when to update, Project overview, and Ownership/contacts. For each file we parse its Markdown structure into heading-content pairs.
A coder reads the heading and the text under it, then assigns the pair to the category that best matches its intent. For example, a section titled ``Running tests'' and one titled ``How to test'' both map to the Test/CI commands category, even though their headings differ. If a file has multiple sections that express the same category, the category is recorded once for that file. Two authors independently classified an agreement sample of 400 instruction files using the 16-category taxonomy and a shared coding guide. The guide included category definitions and representative examples. We computed Cohen's kappa \cite{Cohen1960Kappa} on the independently assigned labels and obtained $\kappa = 0.80$. The authors discussed disagreements until they reached consensus on the category assignments and coding guidelines. Following this reconciliation, one author classified the remaining instruction files using the agreed guidelines. We subsequently used the resulting labels to calculate category prevalence across the four instruction formats.\\
\paragraph{\textbf{Statistical Tests.}}
To identify statistically significant differences across the four file formats, we use the Mann-Whitney U test at a significance level of $\alpha = 0.05$. For comparisons across more than two groups we use the Kruskal-Wallis  \cite{ref1}. We compute Cliff's delta $d$~\citep{CliffsDelta1993} to quantify effect size using the following thresholds:
\begin{equation}
\text{Effect size} =
\begin{cases}
\textit{negligible,} & \text{if } |d| < 0.147 \\
\textit{small,}      & \text{if } 0.147 \leq |d| \leq 0.33 \\
\textit{medium,}     & \text{if } 0.33  < |d| \leq 0.474 \\
\textit{large,}      & \text{if } 0.474 < |d| \leq 1
\end{cases}
\end{equation}
\section{Results}
\label{sec:results}
\subsection{RQ1: How Are Agent Instruction Files Adopted Across Repositories and Formats?} \label{sec:rq1} 
\paragraph{\textbf{Motivation.}}
The motivating example  in section \ref{sec:motivating} shows that a large, actively maintained repository can carry more than one instruction format at the same time. It is not clear whether this multi-format practice is common across projects or limited to a few high-profile cases. Differences in adoption across repository popularity levels may reflect differences in development practices. However, the presence of an instruction file alone does not establish its quality or usefulness. Understanding how often each format appears, and in which kinds of repositories, is the foundation for the content and maintenance analyses that follow.

\paragraph{\textbf{Approach.}}
We measure five metrics for the top and bottom cohorts. \emph{Overall adoption rate} is the percentage of repositories that contain at least one instruction file. \emph{Per-format adoption rate} is the percentage of repositories that contain a file of each specific format. A repository can appear in more than one format count, so these percentages are not mutually exclusive. \emph{Multi-file adoption rate} is the percentage of instruction file bearing repositories that contain two or more formats at the same time. \emph{Co-occurrence pairs} count how many repositories contain each specific combination of two formats (for example, \texttt{AGENTS.md} and \texttt{CLAUDE.md} together). \emph{Language-stratified adoption rate} is the overall adoption rate broken down by primary programming language, to check whether format preferences differ across language communities.
We also fit a logistic regression model using log-transformed star count, log-transformed fork count, cohort membership, and primary programming language as predictors, and report odds ratios to quantify which factors are associated with higher adoption. Finally, we divide all 1{,}000 repositories into four star-count quartiles to study how adoption changes continuously across the popularity range.

\begin{figure*}[t]
  \centering
  \includegraphics[width=\linewidth]{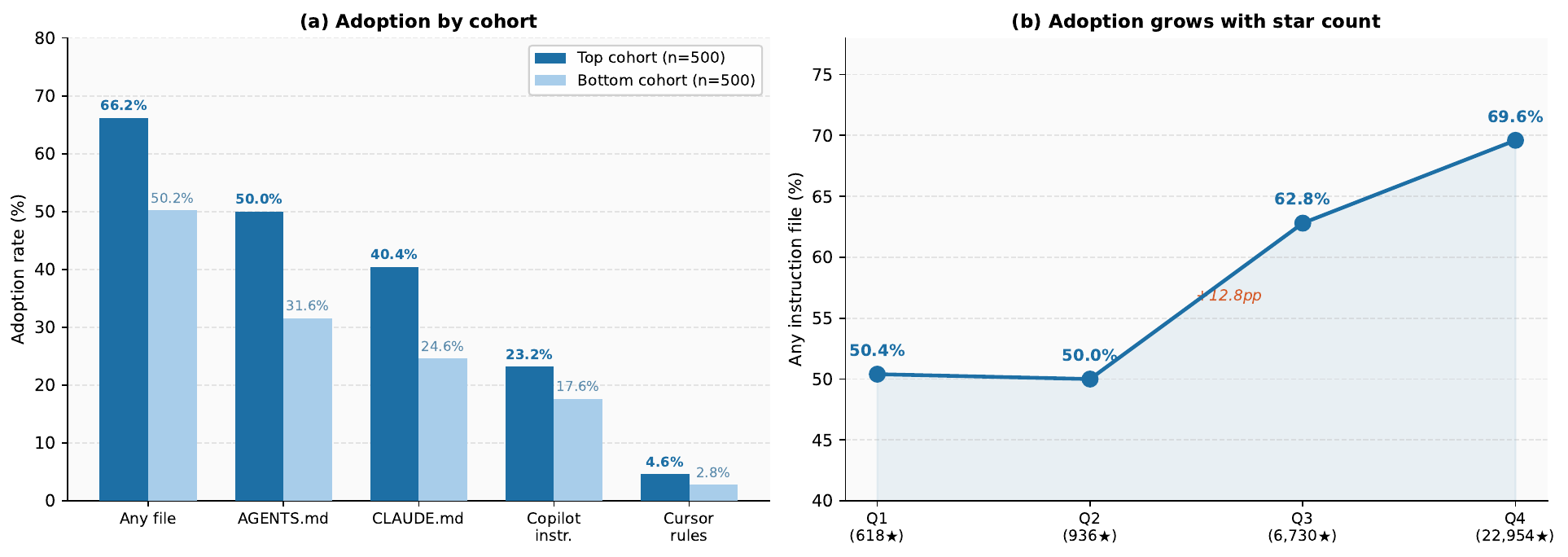}
  \caption{Instruction file adoption.
    (a) Per-format adoption rates in the top cohort (dark blue)
    and bottom cohort (light blue).
    \texttt{AGENTS.md} is the dominant format in both cohorts.
    (b) Overall adoption rate grows with star count,
    accelerating sharply at Q3 (median 6{,}730 stars).}
  \label{fig:rq1}
\end{figure*}

\paragraph{\textbf{Findings.}}
\textbf{Instruction-file adoption is higher among repositories with more stars.} In the higher-popularity cohort, 66.2\% of repositories contain at least one instruction file, compared with 50.2\% in the lower-popularity cohort (Figure~\ref{fig:rq1}a). Across star-count quartiles, adoption rates are 50.4\%, 50.0\%, 62.8\%, and 69.6\%, respectively (Figure~\ref{fig:rq1}b). Adoption is therefore generally higher in the upper quartiles, although the pattern is not strictly monotonic. The logistic regression also indicates a positive association between log-transformed star count and instruction-file presence, with an odds ratio of 1.171. 

\textbf{\texttt{AGENTS.md} is the dominant format, except in C\#.} Across both cohorts, \texttt{AGENTS.md} leads at 50.0\% (top) and 31.6\% (bottom), followed by \texttt{CLAUDE.md} at 40.4\% and 24.6\%. \texttt{copilot-instructions.md} ranks third at 23.2\% and 17.6\%, while \texttt{.cursorrules} remains rare at 4.6\% and 2.8\%. C\# repositories break this pattern: Copilot instructions leads at 53.4\% across both tiers, reflecting the deep integration of GitHub Copilot in the Visual Studio ecosystem~\citep{ConfiguringAgenticTools2026}.

\textbf{Multiple instruction formats coexist more frequently in the higher-popularity cohort.} Among repositories containing at least one instruction file, 42.2\% of the higher-popularity cohort and 21.6\% of the lower-popularity cohort contain two or more formats. The most common co-occurring pair is AGENTS.md and CLAUDE.md, observed in 231 repositories. Such coexistence creates opportunities for guidance to diverge when related instructions are maintained independently. However, the presence of multiple formats does not itself establish inconsistent or outdated guidance.

\begin{tcolorbox}[colback=gray!7, colframe=gray!40, title=\textbf{Answer to $\mathbf{RQ_1}$}] Instruction-file adoption is higher in the upper repository-popularity cohorts. AGENTS.md is the most frequently observed format overall. Among instruction-bearing repositories in the higher-popularity cohort, 42.2\% contain multiple formats. Their coexistence creates potential coordination challenges but does not necessarily indicate inconsistent guidance.
\end{tcolorbox}
\subsection{\texorpdfstring{$\mathbf{RQ_2}$}{RQ2}: How Similar Are Agent Instruction Files in Content?} \label{sec:rq2}
\paragraph{\textbf{Motivation.}}
RQ1 shows that multiple agent instruction
file formats frequently coexist within
the same repository. However, their
coexistence does not reveal whether
they provide similar or different
guidance. As illustrated by the Dawarich
example, two instruction files may
address the same development activity
while prescribing different tools or
workflows. Such differences may reflect
agent-specific requirements or introduce
inconsistencies in repository guidance.
Therefore, we investigate what information
categories are covered across the four
formats and how consistent their
instructions are within the same repository.

\paragraph{\textbf{Approach.}}
We analyse 3{,}387 instruction files: 1{,}677 \texttt{AGENTS.md}, 1{,}107 \texttt{CLAUDE.md}, 538 \texttt{copilot-instructions.md}, and 65 \texttt{.cursorrules}. We study content at two levels (a) Topic-level similarity, (b) Instruction-level consistency.
 
\textbf{(a) Topic-level similarity.} We measure how much sibling files overlap in the subjects they cover using \emph{Jaccard similarity}. Jaccard similarity compares two sets by computing the ratio of shared items to total items: a score of 1.0 means perfect overlap and 0.0 means no shared items. We compute Jaccard at two sub-levels using the raw instruction files directly. \emph{Heading-level Jaccard} compares the literal section titles in each file. \emph{Category-level Jaccard} compares the semantic topics covered by each file, regardless of how those topics are named. For example, ``Running tests'' and ``Test commands'' are different headings but map to the same category. We report both Jaccard values for every co-occurring format pair within the same repository.
 
\textbf{Category labeling and prevalence.} We map instruction sections to the 16-category taxonomy described in Section \ref{sec:taxonomy}. Two authors independently coded the agreement sample, obtaining Cohen's $\kappa = 0.80$. After resolving disagreements and finalizing the coding guidelines, one author classified the remaining files. We use these category assignments to calculate prevalence across the four formats, as reported in Table \ref{tab:categories}.
 
\textbf{(b) Instruction-level consistency.} Topic overlap tells us whether sibling files cover the same ground. It does not tell us whether they give agents the same instructions about that ground. To assess instruction-level consistency, we first construct all eligible section pairs
by matching sections assigned to the same
category across different instruction file
formats within the same repository.
From these pairs, we randomly select
386 for manual examination, following
the conventional sample-size calculation
for a 95\% confidence level with a 5\%
margin of error ~\citep{Cochran1977Sampling}. We additionally examine their applicability scopes and find that 186 pairs (48.2\%) share the same repository-wide or directory scope, 193 (50.0\%) belong to different directory or component scopes, and 7 (1.8\%) have unclear scope. Thus, our comparison captures differences in guidance across formats without assuming that all paired instructions apply to the same component. We embed each section using the all-MiniLM-L6-v2 sentence embedding model~\citep{Reimers2019SentenceBERT}, which converts text into a numerical vector that captures meaning, placing sections with similar meaning close together in vector space. We chose this model because it is the most downloaded and most
liked sentence-similarity model on Hugging Face, indicating strong
community trust and widespread adoption. This model employs
self-attention distillation to capture contextual information effectively while maintaining computational efficiency \cite{backtobasics}. Two annotators independently label each pair as \emph{equivalent} or \emph{non-equivalent}. A pair is \emph{equivalent} if both sections give consistent guidance: either the same operational instruction in different words, or different but compatible information that an agent can follow without conflict. A pair is \emph{non-equivalent} if an agent following one section would behave differently from one following the
other. For example, one section says \texttt{pnpm install} while the other says \texttt{npm install}. Within non-equivalent pairs we further record whether any pair is \emph{contradictory}: giving instructions that an agent cannot simultaneously follow. We report the proportion of equivalent and non-equivalent pairs with 95\% confidence intervals and inter-annotator agreement using Cohen's $\kappa$~\citep{Cohen1960Kappa}.

\begin{table}[ht]
\caption{Instruction-level consistency for 386
section pairs from repositories maintaining two
or more instruction formats.}
\label{tab:consistency}
\centering
\small
\setlength{\tabcolsep}{5pt}
\begin{tabular}{lrr}
\toprule
\textbf{Label} & \textbf{Count} & \textbf{\%} \\
\midrule
\quad Same guidance, different words & 126 & 32.6\% \\
\quad Compatible, complementary & 198 & 51.3\% \\
\textbf{Total equivalent/compatible} & \textbf{324} & \textbf{83.9\%} \\
\midrule
\quad Divergent & 62 & 16.1\% \\
\quad Contradictory & 0 & 0.0\% \\
\textbf{Total non-equivalent} & \textbf{62} & \textbf{16.1\%} \\
\bottomrule
\vspace{-10pt}
\end{tabular}
\end{table}
 
\begin{figure*}[t]
  \centering
  \includegraphics[width=\linewidth]{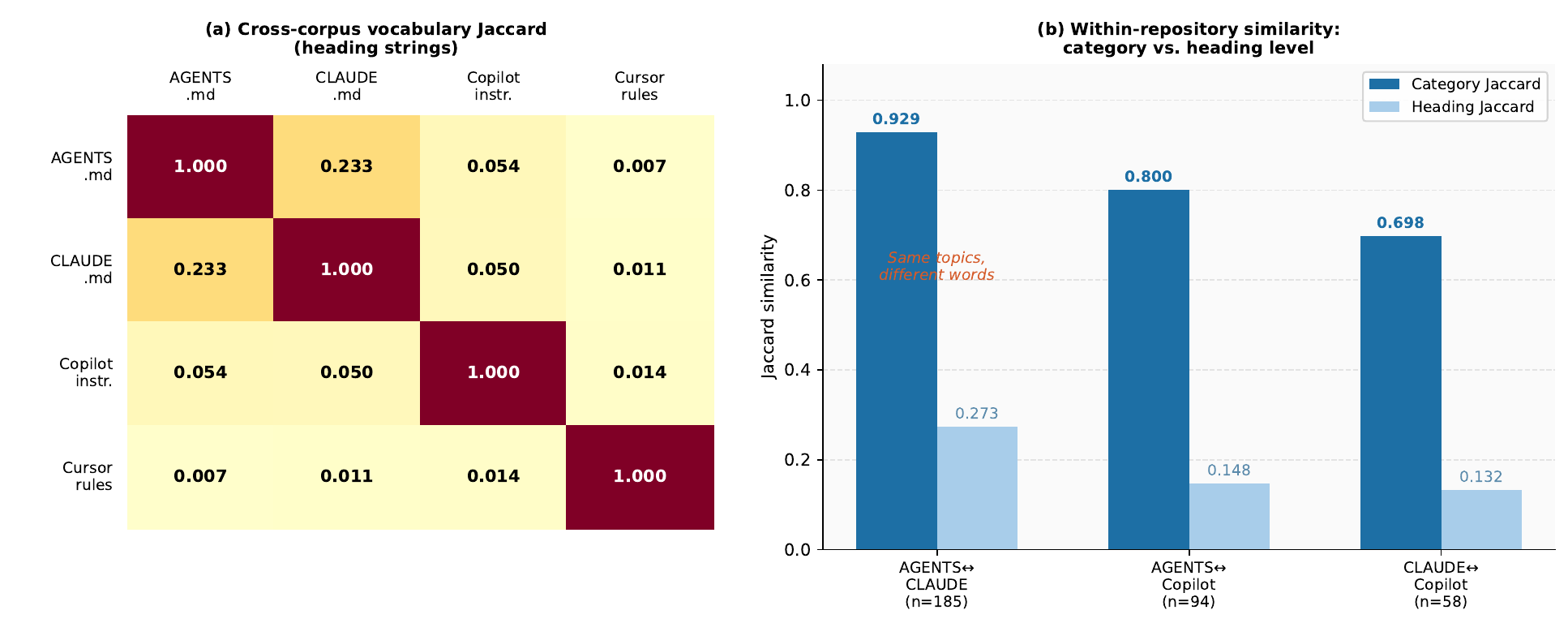}
  \caption{Category similarity across the four formats.
    (a)~Cross-corpus vocabulary Jaccard at the heading-string
    level.
    (b)~Within-repository Jaccard for co-occurring format
    pairs: category-level similarity}
  \label{fig:rq2_jaccard}
\end{figure*}

\paragraph{\textbf{Findings.}}
\textbf{The four formats cover the same topics but describe them
in different words.} Figure~\ref{fig:rq2_jaccard} shows the section similarity results. When \texttt{AGENTS.md} and \texttt{CLAUDE.md} coexist in the same repository, their category-level Jaccard is 0.929 (median), meaning the two files cover nearly identical topics. Their heading-string Jaccard is only 0.273 (median), meaning fewer than one in three section headings uses the same wording. Both files have a section about testing but one calls it ``Running tests'' and the other ``Test commands.'' Both have a section about building but one says ``Setup'' and the other says ``Build and run.'' The files are covering the same ground in different words. This matters because when a developer updates one section, there is no obvious heading-level signal that a differently-named section in a sibling file covers the same topic and may need the same update. \texttt{.cursorrules} is an exception: its vocabulary Jaccard with any other format is below 0.015, reflecting its distinct compact rule syntax that differs structurally from the other three formats~\citep{Jiang2026Cursor, CursorRulesGitHub2026}.
 
\begin{table*}[t]
\caption{Category prevalence across the four instruction
  file formats.
  Prevalence is the percentage of files of each type that
  contain at least one section mapped to that category.
  }
\label{tab:categories}
\centering
\small
\setlength{\tabcolsep}{5pt}
\begin{tabular}{lrrrrr}
\toprule
\textbf{Category} &
\textbf{AGENTS} & \textbf{CLAUDE} &
\textbf{Copilot} & \textbf{Cursor} &
\textbf{Mean\%} \\
\midrule
 Do / Don't rules            & 80.1 & 69.8 & 90.5 & 84.6 & 81.2 \\
Build / run / install       & 87.1 & 71.2 & 70.1 & 55.4 & 71.0 \\
Repository structure        & 77.5 & 68.3 & 67.8 & 69.2 & 70.7 \\
Tooling / agent config      & 76.3 & 81.8 & 61.9 & 36.9 & 64.2 \\
Examples / snippets         & 71.7 & 64.1 & 57.4 & 60.0 & 63.3 \\
Domain models               & 69.5 & 64.0 & 59.9 & 50.8 & 61.0 \\
Test / CI commands          & 79.5 & 62.1 & 57.8 & 36.9 & 59.1 \\
API / interface contracts   & 64.3 & 55.6 & 51.9 & 52.3 & 56.0 \\
Architecture / design       & 60.7 & 58.3 & 55.0 & 44.6 & 54.6 \\
Code style / formatting     & 51.9 & 49.0 & 58.9 & 50.8 & 52.6 \\
Git / PR conventions        & 58.1 & 48.8 & 52.8 & 15.4 & 43.8 \\
Project overview            & 48.6 & 39.5 & 42.8 & 23.1 & 38.5 \\
Performance requirements    & 32.7 & 39.2 & 36.2 & 26.2 & 33.6 \\
Security / secrets          & 36.3 & 29.9 & 32.0 & 21.5 & 29.9 \\
Maintenance / when to update& 36.6 & 39.7 & 26.8 &  3.1 & 26.6 \\
Ownership / contacts        & 25.6 & 17.9 & 16.2 & 15.4 & 18.8 \\
\midrule
\textit{Total Number of Files}
   & \textit{1,677} & \textit{1,107}
   & \textit{538}   & \textit{65}
   & \textit{3,387} \\
\bottomrule
\end{tabular}
\end{table*}

\textbf{A subset of corresponding sections provides divergent guidance.} Table~\ref{tab:consistency} presents the instruction-level consistency results for 386 examined section pairs. Inter-annotator agreement is Cohen's $\kappa = 0.85$. Of these pairs, 324 (83.9\%) provide equivalent or complementary guidance: 126 (32.6\%) express the same instructions in different words, while 198 (51.3\%) provide compatible, non-conflicting information. The remaining 62 pairs (16.1\%) provide divergent guidance that may lead to different agent behavior depending on which file is consulted. No explicitly contradictory pair was identified. These results indicate that differences between corresponding sections are not necessarily direct conflicts. Some differences may reflect agent-specific requirements or different levels of detail, while others may arise from independently maintained guidance. The high proportion of compatible but non-identical sections suggests that consolidating instruction files requires attention to complementary information. 
\textbf{The 16 section categories vary widely in how often they appear across the four formats.} Table~\ref{tab:categories} shows the prevalence of each category across all four formats. The most common categories are Do/Don't rules (81.2\% mean), build and install commands (71.0\%), and repository structure (70.7\%). These categories appear in most files across every format, suggesting that developers across all four tool communities independently converge on the same set of core topics. At the other end, ownership information appears in only 18.8\% of files on average, maintenance guidance in 26.6\%, security rules in 29.9\%, performance requirements in 33.6\%, and project overview in 38.5\%. These five categories are absent from most files of every format. Their consistent absence across all formats and all cohort tiers confirms the finding of Chatlatanagulchai et al. ~\cite{Chatlatanagulchai2025AgentREADMEs} and extends it to all four formats.

\begin{tcolorbox}[colback=gray!7, colframe=gray!40,
  title=\textbf{Answer to $\mathbf{RQ_2}$}]
Instruction files from different formats frequently cover overlapping topics, although their headings differ. For AGENTS.md--CLAUDE.md pairs, median category level Jaccard similarity is 0.929, compared with 0.273 at the heading level. Among 386 examined section pairs, 83.9\% provide equivalent or complementary  while 16.1\% are non-equivalent. Security, performance, maintenance, project overview, and ownership are among the least frequently represented categories.
\end{tcolorbox}
\subsection{\texorpdfstring{$\mathbf{RQ_3}$}{RQ3}: How Do Agent Instruction Files Evolve Over Time?} \label{sec:rq3}
\paragraph{\textbf{Motivation.}}
RQ1 and RQ2 characterize instruction files at the inspected repository revision. RQ3 examines their recorded modification histories. Instruction files may require updates when repository practices, build procedures, or development conventions change. However, infrequent modification does not necessarily indicate outdated content, and deletion does not necessarily indicate that the removed guidance lacked value. We therefore examine modification frequency, ownership, churn, update associations, and category-level additions and deletions to characterize observable maintenance practices.

\paragraph{\textbf{Approach.}}
We analyze 5,396 heading-level edit events across 448 unique instruction files from 283 of our 1,000 selected repositories, covering 943 pull requests recorded in the AIDev dataset~\citep{Li2026AIDev}. We timestamp each event using the pull request's \texttt{merged\_at} value, falling back to \texttt{created\_at} for unmerged pull requests. To characterize instruction-file maintenance, we examine five dimensions: modification frequency (touch count), contributor involvement (ownership concentration), extent of content removal (churn ratio), temporal evolution (lifecycle pattern), and update context (update trigger). \textbf{Touch count} is the number of recorded modification events associated with an instruction file in the available history. A touch count of one indicates that only one event was observed. \textbf{Ownership concentration} is the fraction of all touch events made by the single most active contributor. A score of 1.0 means one person made every change. \textbf{Churn ratio} is the number of lines deleted divided by the total lines changed in a typical update. A churn ratio of 0.0 means updates only add content. A churn ratio of 0.4 means that 40\% of changed content in a typical update is deleted, indicating a higher proportion of
content removal relative to
addition-oriented updates. \textbf{Lifecycle pattern} characterizes the distribution of recorded modification events. We distinguish single-touch files, sparse updates (occasional updates with no pattern), burst-taper patterns (intensive early updates then inactivity), and continuous updates (updates spread across the observation period). These labels describe observed histories rather than the long-term status or usefulness of the instruction files. \textbf{Update trigger} classifies what each touch event is associated with. A proactive update has no obvious external cause: a developer updated the file independently. A proactive co-change means the instruction file was modified in the same pull request as code files. A reactive update follows a failed or heavily reviewed agentic pull request within a 14-day lookback window.

\textbf{Category-level heading changes.} To characterize how instruction topics change, we extract heading additions and deletions from instruction-file patches in the AIDev pull request history. We identify added and removed Markdown headings using the corresponding added and deleted lines in each patch and map these headings to the 16 semantic categories. We define the category addition ratio (CAR) for category $c$ as:
\begin{equation}
CAR = \frac{A_c}{A_c + D_c},
\end{equation}
where $A_c$ and $D_c$ denote the number of recorded category additions and deletions, respectively. A higher ratio indicates that additions account for a greater proportion of the observed category edits. CAR and churn ratio capture different aspects of instruction-file modifications. Churn ratio measures the proportion of changed lines that are deletions, reflecting the extent of content removal during file updates. In contrast, CAR measures the proportion of added headings among all added and deleted headings mapped to a particular instruction category. Thus, churn ratio operates at the line level, whereas CAR captures heading-level addition and deletion patterns across semantic categories. A file may undergo substantial content modifications without changes to its headings, which would affect churn ratio but not CAR.

We compute churn ratio, ownership concentration, and co-change rate following the build coupling and build ownership metrics~\cite{McIntosh2011Build}. Lifecycle pattern classification follows the README lifecycle taxonomy~\citep{Gaughan2025README}. We compare formats using Kruskal-Wallis tests with Cliff's delta for pairwise effect sizes~\citep{ref1,CliffsDelta1993}. 

\begin{table}[t]
\caption{Category additions, deletions,
and category addition ratios (CAR) across
instruction categories.}
\label{tab:durability}
\centering
\small
\setlength{\tabcolsep}{4pt}
\begin{tabular}{lrrrr}
\toprule
\textbf{Category} &
\textbf{Additions} & \textbf{Deletions} & \textbf{Total changes} &
\textbf{CAR} \\
\midrule
\multicolumn{5}{l}{\textit{CAR $\geq$ 0.80}} \\
\midrule
Ownership / contacts          &    7 &   0 &    7 & 1.000 \\
Maintenance / when to update  &    7 &   1 &    8 & 0.875 \\
Security / secrets            &  116 &  18 &  134 & 0.866 \\
Git / PR conventions          &  309 &  51 &  360 & 0.858 \\
Do / Don't rules              &  685 & 145 &  830 & 0.825 \\
Test / CI commands            &  266 &  58 &  324 & 0.821 \\
Repository structure          &  336 &  78 &  414 & 0.812 \\
API / interface contracts     &  135 &  33 &  168 & 0.804 \\
Build / run / install         &  959 & 239 & 1{,}198 & 0.801 \\
\midrule
\multicolumn{5}{l}{\textit{CAR $<$ 0.80}} \\
\midrule
Performance requirements      &  131 &  34 &  165 & 0.794 \\
Examples / snippets           &   73 &  19 &   92 & 0.793 \\
Architecture / design         &  329 &  87 &  416 & 0.791 \\
Code style / formatting       &  215 &  57 &  272 & 0.790 \\
Project overview              &  190 &  66 &  256 & 0.742 \\
Tooling / agent config        &  383 & 149 &  532 & 0.720 \\
Domain models                 &  156 &  64 &  220 & 0.709 \\
\bottomrule
\end{tabular}
\end{table}

\paragraph{\textbf{Findings.}}
\textbf{Most files have few modification events and are maintained by a single person.} The median touch count is 2.0 across all four formats: the typical file has exactly two recorded modification events. The median author count is 1.0 for every format. The mean ownership share of the top author is 0.962 for \texttt{AGENTS.md}, 0.910 for \texttt{CLAUDE.md}, 0.908 for \texttt{.cursorrules}, and 0.885 for Copilot instructions. Ownership concentration differs significantly across formats (Kruskal-Wallis $H = 12.48$, $p = 0.006$), with \texttt{AGENTS.md} showing the highest concentration (Cliff's $d = 0.167$ vs.\ Copilot instructions, small effect). Ownership concentration may affect how instruction files are maintained, particularly when a repository contains several formats. If different contributors maintain related files, coordinating updates may require additional effort. However, our analysis does not establish that ownership concentration or separation caused the divergent guidance observed in RQ2.

\textbf{Four in ten files have only one recorded modification event.} Figure~4a presents the distribution of observed life-cycle patterns. Overall, 40.4\% of files in the maintenance subset have a single recorded touch event. Copilot instruction files have the highest single-touch proportion (47.7\%), while CLAUDE.md has the lowest (34.4\%) and the highest proportion classified as continuously updated (32.0\%). A single observed event does not establish that a file became outdated, or remained unchanged throughout its entire existence. Nevertheless, the distribution illustrates variation in recorded maintenance activity across formats.

\begin{figure*}[ht]
  \centering
  \includegraphics[width=\linewidth]{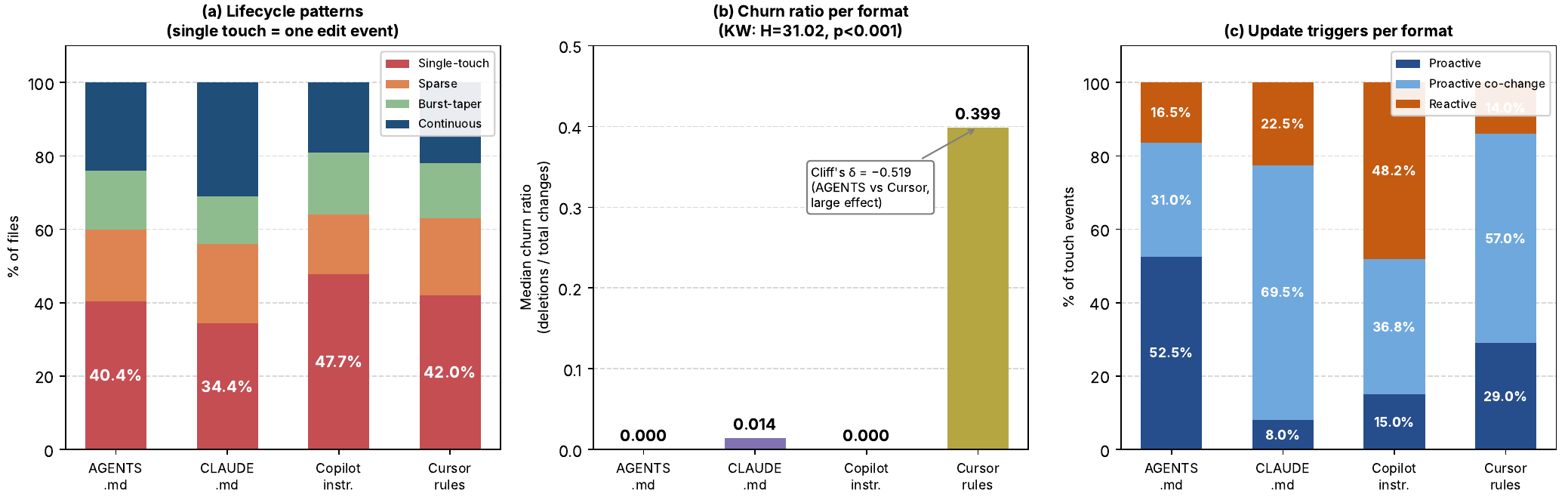}
  \caption{Maintenance patterns across the four formats.
    (a)~Lifecycle distribution.
    (b)~Churn ratio.
    (c)~Update triggers.}
  \label{fig:rq3_maintenance}
\end{figure*}

\textbf{The four formats exhibit different observed modification patterns.} Figure~\ref{fig:rq3_maintenance}b shows the churn ratio distribution. Churn ratio differs significantly across formats (Kruskal-Wallis $H = 31.02$, $p < 0.001$). The median churn ratios indicate different patterns in the observed changes. AGENTS.md and Copilot instructions have median ratios of 0.0, while CLAUDE.md has a median of 0.014 and .cursorrules has a median of 0.399. These results suggest that observed  AGENTS.md modifications are more addition-oriented, whereas cursorrules modifications involve a higher proportion of deleted lines.
 
\textbf{Each format is associated with different update triggers.} Figure~\ref{fig:rq3_maintenance}c shows the trigger distribution. \texttt{AGENTS.md} is predominantly updated proactively and independently: 52.5\% of its modifications are standalone proactive updates, not associated with any code change or agent failure. \texttt{CLAUDE.md} is almost always modified alongside code changes: 65.2\% of its updates are proactive co-changes, and its co-change rate with non-instruction files is 0.845. Copilot instructions is predominantly updated reactively: 48.2\% of its updates are associated with a failed or heavily reviewed
agentic pull request. \texttt{.cursorrules} falls between these with 54.2\% proactive co-change. Feature pull requests are the most common associated event across all formats (33.9\% of all 5{,}396 events).
 
These update associations vary across
instruction formats. CLAUDE.md updates
more frequently coincide with code
changes, whereas Copilot instruction
updates are more frequently associated
with recent failed or heavily reviewed
agentic pull requests. Such associations
may indicate differences in maintenance
practices, but do not establish the
developers' motivations for updating
these files.
 
\textbf{Heading-level changes are predominantly additions rather than deletions.} Table~\ref{tab:durability} presents the numbers of added and deleted headings associated with each instruction category. Build/run/install has the highest number of heading additions (959), followed by Do/Don't rules (685) and Tooling/agent config (383). Ownership/contacts has a CAR of 1.000, indicating that all seven recorded heading-change events associated with this category were additions. Maintenance/when to update (0.875), Security/secrets (0.866), and Git/PR conventions (0.858) also exhibit high proportions of heading additions. In contrast, Domain models (0.709), Tooling/agent config (0.720), and Project overview (0.742) have relatively higher proportions of heading deletions, although additions still outnumber deletions. These findings suggest that observed heading-level changes are generally addition-oriented across all 16 categories. However, these ratios do not establish whether individual categories or their associated instructions remain unchanged over time.

\begin{tcolorbox}[colback=gray!7, colframe=gray!40,
  title=\textbf{Answer to $\mathbf{RQ_3}$}]
Four in ten instruction files have only a single modification event (Figure~\ref{fig:rq3_maintenance}a). The four formats are updated in structurally different ways: \texttt{AGENTS.md} grows by accumulation, \texttt{CLAUDE.md} co-evolves with code, Copilot instructions is updated reactively after agent failures, and \texttt{.cursorrules} exhibits the highest proportion of deleted lines, with updates primarily occurring alongside code changes (Figure~\ref{fig:rq3_maintenance}b, \ref{fig:rq3_maintenance}c). Category-level changes are predominantly additions rather than deletions, particularly for ownership, security, and build instructions, while tooling configuration and domain models show relatively higher proportions of deletions (Table~\ref{tab:durability}).

\end{tcolorbox}

\section{An Evidence-Based Template for Agent Instruction Files} \label{sec:template}

Our empirical findings reveal both recurring patterns and differences in the content and maintenance of agent instruction files. Building on these observations, we derive an evidence-based template that organizes repository guidance into a consistent structure while accommodating project-specific requirements.

Prior research on repository documentation demonstrates the value of systematically characterizing documentation content \citep{Prana2019README}. Recent studies of agentic tool configurations similarly show that repository-level context files are an important mechanism for communicating
development guidance \citep{Chatlatanagulchai2025AgentREADMEs, ConfiguringAgenticTools2026}. However, the presence of instructions does not necessarily establish their usefulness for downstream coding tasks \citep{Gloaguen2026Evaluating}.

Accordingly, our template is intended as an empirically motivated authoring aid, rather than a universally optimal instruction structure. It contains 16 categories organized into three tiers: ten Required, four Gap, and two Recommended.

\subsection{Deriving the Template from Empirical Evidence} \label{sec:template_derivation}
We adopt the 16-category taxonomy used in RQ2~\citep{Chatlatanagulchai2025AgentREADMEs} and combine two measurements to determine category tiers: prevalence and the category addition ratio (CAR).

\paragraph{Category prevalence.} We measure the proportion of instruction files containing each category separately for the four formats. For category $c$, the mean prevalence is:

\begin{equation}
\overline{P}_c =
\frac{1}{4}\sum_{f=1}^{4}P_{c,f},
\end{equation}

where $P_{c,f}$ denotes the prevalence of category $c$ in format $f$. We use an unweighted mean to prevent the larger format-specific corpora from dominating the tier assignment. Consequently, a mean prevalence of 50\% does not imply that the category appears in at least half of the files within each individual format.

\paragraph{Category addition ratio.} We use the heading-change events examined in RQ3 to characterize the balance of recorded additions and deletions associated with each category. For category $c$:

\begin{equation}
CAR_c = \frac{A_c}{A_c+D_c},
\end{equation}

where $A_c$ and $D_c$ denote the numbers of recorded heading additions and deletions mapped to category $c$, respectively. CAR measures the proportion of these events that are additions. It does not measure section survival, content persistence, or the correctness of the guidance.

\paragraph{Tier assignment.}
We combine the two measurements using the following selection policy:

\begin{equation}
\operatorname{Tier}(c)=
\begin{cases}
\text{Required}, & \overline{P}_c \geq 50\%,\\
\text{Gap}, & \overline{P}_c < 50\%
              \land CAR_c \geq 0.80,\\
\text{Recommended}, & \text{otherwise}.
\end{cases}
\end{equation}
Categories with mean prevalence of at least 50\% form the Required tier. Among the remaining categories, those with CAR values of at least 0.80 form the Gap tier. All other categories are classified as Recommended.These thresholds provide a transparent and reproducible selection policy. They are heuristic design decisions informed by the observed corpus, rather than statistically optimized cutoffs or evidence that individual categories improve agent performance.

\subsection{Template Categories and Authoring Guidance} \label{sec:template_categories}
Table~\ref{tab:template} presents the 16 categories, their empirical measurements, tier assignments, and corresponding authoring guidance. The guidance specifies the kinds of information developers should document without prescribing unverified commands, contacts, or project-specific constraints.

\begin{table*}[t]
\caption{Evidence-based agent instruction
file template. Prevalence is the unweighted
mean across four formats; CAR denotes the
category addition ratio.}
\label{tab:template}
\centering
\small
\setlength{\tabcolsep}{3pt}
\begin{tabular}{
p{0.23\textwidth}lrrp{0.43\textwidth}}
\toprule
\textbf{Category} & \textbf{Tier}
& \textbf{Prev.\%} & \textbf{CAR}
& \textbf{Information to record} \\
\midrule
Do / Don't rules & Required & 81.2 & 0.825
& Required and prohibited agent actions,
including change boundaries. \\
Build / run / install & Required & 71.0 & 0.801
& Verified setup, build, and execution
commands and prerequisites. \\
Repository structure & Required & 70.7 & 0.812
& Important directories, file locations,
and their responsibilities. \\
Tooling / agent configuration & Required
& 64.2 & 0.720
& Agent settings, applicable tools,
permissions, and context references. \\
Examples / snippets & Required & 63.3 & 0.793
& Representative examples of established
project practices. \\
Domain models & Required & 61.0 & 0.709
& Important domain entities, data
structures, and relationships. \\
Test / CI commands & Required & 59.1 & 0.821
& Verified test commands, CI checks,
and validation requirements. \\
API / interface contracts & Required
& 56.0 & 0.804
& Interface constraints, compatibility
requirements, and API change rules. \\
Architecture / design & Required
& 54.6 & 0.791
& Component boundaries, dependencies,
and relevant design constraints. \\
Code style / formatting & Required
& 52.6 & 0.790
& Coding conventions, naming rules,
and formatting tools. \\
\midrule
Security / secrets & Gap & 29.9 & 0.866
& Sensitive-data restrictions and
verified security practices. \\
Maintenance / when to update & Gap
& 26.6 & 0.875
& Conditions triggering instruction
review and update procedures. \\
Ownership / contacts & Gap & 18.8 & 1.000
& Responsible teams and verified
contact or escalation routes. \\
Git / PR conventions & Gap & 43.8 & 0.858
& Branching, commit, review, and
pull-request conventions. \\
\midrule
Project overview & Recommended
& 38.5 & 0.742
& Project purpose and relevant context. \\
Performance requirements & Recommended
& 33.6 & 0.794
& Established performance constraints
and measurement procedures. \\
\bottomrule
\end{tabular}
\end{table*}

\paragraph{\textbf{Required categories.}} The ten Required categories have mean prevalence values of at least 50\%. They cover repository organization, development procedures, coding conventions, interface constraints, and agent-specific configuration. These categories provide the baseline structure of the template.

Here, Required denotes the template's default coverage tier, not an obligation to fabricate information for every category. Developers should include applicable guidance and omit or explicitly mark information that cannot be established from the repository.

\paragraph{\textbf{Gap categories.}}
Four categories exhibit lower prevalence but addition-dominated category-change histories: Security/secrets, Maintenance/when to update, Ownership/contacts, and Git/PR conventions. Their mean prevalence ranges from 18.8\% to 43.8\%, while Category Addition Ratio (CAR) ranges from 0.858 to 1.000. These categories highlight guidance that is less frequently represented in the analyzed files despite a relatively high share of recorded category additions. However, the underlying event counts vary considerably. Ownership/contacts has seven additions and no deletions, while Maintenance/when to update has seven additions and one deletion. Security/secrets has 116 additions and 18 deletions, and Git/PR conventions has 309 additions and 51 deletions. We therefore interpret these ratios as descriptive evidence rather than proof of long-term category importance.

\paragraph{\textbf{Recommended categories.}} Project overview and Performance requirements have mean prevalence below 50\% and CAR below 0.80. Their Recommended classification does not imply that they are unimportant. For example, repositories with explicit latency, resource, or throughput constraints should document those requirements.

More generally, the tiers describe empirical patterns and support authoring decisions; they do not rank the universal importance of individual categories. This distinction is particularly relevant given recent evidence that unnecessary repository context can increase coding-agent cost without improving task success~\citep{Gloaguen2026Evaluating}.

\subsection{Maintenance Protocol} \label{sec:maintenance}
The template also incorporates a maintenance protocol derived from the change patterns observed in RQ3. We identify three review triggers: agent-related errors, repository changes, and periodic accuracy checks. These are practical recommendations informed by our observations rather than experimentally validated interventions.

\paragraph{Agent-related errors.} Reactive associations account for 48.2\% of Copilot instruction modifications. When an agent error or code review reveals incorrect, missing, or outdated guidance, maintainers should examine the relevant instruction category. However, an agent failure should not automatically result in a new rule without first establishing whether the instructions contributed to the problem.

\paragraph{Repository changes.} Proactive co-changes account for 65.2\% of \texttt{CLAUDE.md} modifications. Changes affecting build commands, testing procedures, directory structure, dependencies, or interface requirements should therefore prompt a review of related instruction categories. Where multiple agent instruction files describe the same practice, maintainers should check the affected representations for consistency.

\paragraph{Periodic accuracy checks.} In the maintenance subset, 40.4\% of instruction files have only one recorded modification event. This does not demonstrate that the files are outdated, but motivates periodic verification of documented commands, paths, constraints, and contacts. The review frequency should depend on repository activity; our observations do not establish an optimal interval.

Developers can record these review triggers under the Maintenance/when to update category and identify responsible reviewers under Ownership/contacts. Together, these categories make instruction maintenance an explicit part of the proposed authoring framework.

\section{AGIT-Instruct: A Benchmark for Template-Based Instruction Generation} \label{sec:generation}
The proposed template identifies the information that agent instruction files should contain. However, specifying the expected categories does not guarantee that automatically generated instructions preserve developers' intended guidance. To investigate this challenge, we introduce \textbf{AGIT-Instruct}, a benchmark for evaluating automated agent instruction-file generation. We demonstrate its application through template-guided generation using four configurations and examine both structural coverage and content alignment. We also compare generation with and without template guidance to explore the template's practical applicability.

\subsection{AGIT-Instruct Benchmark}
\label{sec:agit_benchmark}

Existing software engineering benchmarks evaluate language models on tasks such as code generation and repository-level issue resolution~\cite{codegen1, Jimenez2024SWEbench}. AGIT-Instruct complements these efforts by focusing on the generation of agent instruction files and their correspondence with developer-written guidance.

\paragraph{\textbf{Generic instruction intents.}} Agent instruction files often contain repository-specific paths, commands, and identifiers. To facilitate comparison across repositories, we use GPT-4o-mini to transform instruction sections into generic intents. For example, an instruction to execute a specific test command before committing is generalized into a requirement to run the project's tests before committing. We apply this transformation to both developer-written references and generated instructions before automatic content evaluation. This reduces dependence on repository-specific wording, but may also abstract away operational details. Consequently, AGIT-Instruct evaluates alignment with generalized developer intent rather than directly verifying the executability or repository-specific correctness of generated instructions.

\paragraph{\textbf{Dataset construction.}}
We construct AGIT-Instruct from the four
instruction formats examined in our empirical
study: \texttt{AGENTS.md}, \texttt{CLAUDE.md},
GitHub Copilot instructions, and Cursor rules.
We retain files containing at least two
instruction sections to support category-level
evaluation. This initial filtering yields
2,853 files from 552 repositories, comprising
1,565 \texttt{AGENTS.md}, 819 \texttt{CLAUDE.md},
447 Copilot instruction files, and 22 Cursor
rule files. We map the instruction sections to the
16-category taxonomy used in our empirical
analysis and apply generic-intent filtering
to prepare the benchmark. To avoid potential
data leakage, we partition the data at the
repository level, ensuring that instruction
files from the same repository do not appear
in different partitions. We construct the evaluation set using
category-coverage criteria rather than
simple random sampling. Specifically,
we select 62 instruction files such that
each of the 16 categories is represented
by at least 25 developer-written reference
instances. This selection ensures sufficient
representation of less frequent categories
and enables category-level evaluation.
A category instance denotes the presence
of a particular instruction category
within a file, counted once per file.

Following filtering and repository-disjoint
partitioning, the resulting benchmark
contains 1,815 files from 522 repositories:
1,432 training files, 321 validation files,
and 62 evaluation files. The evaluation
partition contains 556 developer-written
reference category instances. The training
partition is used for retrieval and model
adaptation, the validation partition for
model development, and the evaluation
partition for comparing generation
configurations. Because the evaluation set was selected
to ensure minimum representation of
each category, its distribution may
differ from that of naturally occurring
instruction files. We therefore
interpret the results as performance
on a category-coverage-oriented
benchmark rather than as an estimate
of performance across the entire
repository population.

\paragraph{\textbf{Benchmark dimensions.}}
AGIT-Instruct supports two complementary evaluation dimensions. Structural coverage measures whether generated instructions represent the expected categories, while content alignment measures their correspondence with developer-written references. This separation is important because generating a category heading does not necessarily imply that the associated instructions are complete or correct.

\subsection{Generation Approaches} \label{sec:generation_approaches}

We use the proposed template as explicit generation guidance. The template organizes 16 categories into ten Required, four Gap, and two Recommended categories. We evaluate four generation configurations on the same test set. All configurations receive the target instruction-file format, programming language, and repository popularity information, along with a short repository summary.

\paragraph{\textbf{GPT-4o-mini zero-shot.}} The model receives a category checklist derived from the template, including guidance on missing content, and generates instructions without retrieved examples. Here, zero-shot refers to the absence of examples rather than the absence of template guidance.

\paragraph{\textbf{GPT-4o-mini RAG.}} Following retrieval-augmented generation \citep{Lewis2020RAG}, we augment the template-guided prompt with three examples retrieved from the repository-disjoint training partition. We use \texttt{all-MiniLM-L6-v2} to encode retrieval inputs and rank candidates by cosine similarity. The encoder builds on Sentence-BERT \citep{Reimers2019SentenceBERT}, and similar retrieval configurations have been used in prior code generation research \citep{codegen1,codegen2}.

\paragraph{\textbf{CodeLlama-13B Base.}} We evaluate CodeLlama-13B without task-specific fine-tuning \citep{Roziere2023CodeLlama}, using template-guided prompts and greedy decoding with a maximum of 1,536 output tokens.

\paragraph{\textbf{CodeLlama-13B QLoRA.}} We adapt CodeLlama-13B using parameter-efficient fine-tuning with QLoRA \citep{Hu2022LoRA,Dettmers2023QLoRA}. We use 4-bit NF4 quantization, LoRA rank 8, scaling factor 16, and dropout 0.05. Training uses AdamW with a learning rate of $10^{-4}$ for five epochs. Generation uses greedy decoding with a maximum of 1,536 output tokens.

These configurations illustrate different approaches to generating instruction content. Our objective is not to establish a comprehensive model ranking, but to demonstrate how the proposed template and AGIT-Instruct support systematic evaluation. Complete prompts, fine-tuning configurations, and implementation details are available in our replication package.

\subsection{Evaluation} \label{sec:generation_evaluation}
\paragraph{Structural coverage.}
We measure the presence of each template category in generated instruction files. For a generated file $i$, let $C_i$ be its detected category set and $T$ be the template category set. We define:

\begin{equation}
\operatorname{Coverage}(i,T)=
\frac{|C_i \cap T|}{|T|}.
\end{equation}

We average coverage across the evaluation files and report results for the complete template and each category tier. We additionally report per-category presence, defined as the fraction of the 62 generated files containing the corresponding category. Structural coverage measures category presence, not whether the generated content is consistent with the reference.

\paragraph{Content alignment.} We evaluate the generalized
instruction intents using ROUGE-1 F1 for lexical overlap \citep{Lin2004ROUGE} and cosine similarity using \texttt{all-MiniLM-L6-v2}
embeddings \citep{Reimers2019SentenceBERT}. The use of this encoder for semantic evaluation is also supported by prior research on retrieval-augmented code review, which employs \texttt{all-MiniLM-L6-v2} to compare natural-language technical explanations \citep{contextCodeReview}.

We report two evaluation settings. Full-file alignment compares
the complete generated and reference intent files. Overlap-only alignment compares categories represented in both files. 

\paragraph{Manual evaluation.}
Automatic metrics cannot fully determine whether generated instructions preserve developer requirements. We therefore manually compare RAG-generated category instances against their corresponding developer-written references, targeting at least ten instances from each of the 16 instruction categories. We use four labels: \emph{Match} when the reference intent is fully preserved; \emph{Partial Match} when the main intent is captured but some requirements are omitted or weakened; \emph{Wrong} when the generated guidance is incompatible with the reference; and \emph{Missing} when the expected instruction content is absent.

\subsection{Findings} \label{sec:generation_findings}

\paragraph{\textbf{Application of Template Guidance.}} \label{sec:template_application}

We first examine the practical application of the template by comparing generation with and without explicit category guidance on the AGIT evaluation set. Table~\ref{tab:template_guidance} presents the structural coverage across the three template tiers.

\begin{table}[t]
\caption{Structural category coverage
with and without template guidance. Values represent
mean category presence within each
tier.}
\label{tab:template_guidance}
\centering
\small
\begin{tabular}{lrr}
\toprule
\textbf{Template tier}
& \textbf{Without Template}
& \textbf{With Template} \\
\midrule
Required (10)
& 0.168 & 0.737 \\
Gap (4)
& 0.080 & 0.977 \\
Recommended (2)
& 0.500 & 0.990 \\
\midrule
All categories (16)
& 0.187 & 0.828 \\
\bottomrule
\end{tabular}
\end{table}

The generated files exhibit substantially broader structural coverage with template guidance. Overall coverage increases from 18.7\% to 82.8\%, corresponding to an average of 3.00 and 13.25 represented categories per file, respectively. The Gap tier shows a particularly large difference, increasing from 8.0\% to 97.7\%. These findings suggest that explicit category guidance is associated with broader coverage of the information identified by our empirical template. However, this comparison is exploratory. Both conditions used GPT-4o-mini, the same
base prompt, and identical repository inputs.
The only difference was the inclusion of
explicit template guidance. Tables~\ref{tab:template_guidance}
and~\ref{tab:generation_category_coverage}
report results from the same evaluation
set. 

\paragraph{\textbf{Structural Coverage Across
Generation Configurations.}}
\label{sec:structural_generation_results}
We next examine how consistently the four template-guided configurations represent the 16 categories. Table~\ref{tab:generation_category_coverage} reports category presence across the common 62-file evaluation set.

\begin{table*}[t]
\caption{Category-level structural coverage of four template-guided generation configurations. Values represent the fraction of generated files containing each category. ZS denotes zero-shot; CL denotes CodeLlama-13B.}
\label{tab:generation_category_coverage}
\centering
\scriptsize
\setlength{\tabcolsep}{5pt}
\begin{tabular}{llrrrr}
\toprule
\textbf{Category}
& \textbf{Tier}
& \textbf{ZS}
& \textbf{RAG}
& \textbf{CL Base}
& \textbf{CL QLoRA} \\
\midrule
Repository structure
& Required & 0.55 & 0.55 & 0.45 & 0.16 \\
Build / run / install
& Required & 0.90 & 0.82 & 0.73 & 0.31 \\
Test / CI commands
& Required & 0.95 & 0.90 & 0.79 & 0.02 \\
Code style / formatting
& Required & 0.82 & 0.79 & 0.65 & 0.31 \\
Architecture / design
& Required & 0.77 & 0.68 & 0.55 & 0.45 \\
Domain models
& Required & 0.31 & 0.31 & 0.21 & 0.06 \\
API / interface contracts
& Required & 0.77 & 0.73 & 0.61 & 0.13 \\
Do / Don't rules
& Required & 0.85 & 0.82 & 0.65 & 0.24 \\
Tooling / agent configuration
& Required & 0.84 & 0.71 & 0.71 & 0.29 \\
Examples / snippets
& Required & 0.61 & 0.61 & 0.45 & 0.05 \\
\midrule
Git / PR conventions
& Gap & 0.95 & 0.90 & 0.84 & 0.03 \\
Security / secrets
& Gap & 0.98 & 0.97 & 0.74 & 0.13 \\
Ownership / contacts
& Gap & 1.00 & 0.97 & 0.73 & 0.02 \\
Maintenance / when to update
& Gap & 0.98 & 0.89 & 0.71 & 0.00 \\
\midrule
Project overview
& Recommended & 1.00 & 0.97 & 0.79 & 0.37 \\
Performance requirements
& Recommended & 0.98 & 0.97 & 0.73 & 0.10 \\
\bottomrule
\end{tabular}
\end{table*}


Across the 16 categories, GPT-4o-mini zero-shot achieves the highest structural coverage (0.828), followed by RAG (0.786), CodeLlama Base (0.645), and CodeLlama QLoRA (0.166). Both GPT configurations represent most Gap and Recommended categories, although their Required-category coverage is lower.

Category-level results reveal additional variation. Domain models are represented in only 31\% of files under both GPT configurations, whereas Test/CI commands appear in 95\% of zero-shot and 90\% of RAG outputs. Thus, explicit template guidance does not produce uniform category coverage. A particularly notable result concerns Ownership/contacts. This category is present in 100\% of zero-shot and 97\% of RAG outputs. This indicates that GPT-based generation can represent the ownership category structurally. However, category presence does not establish whether the generated instructions identify appropriate owners or provide actionable contact information.

CodeLlama QLoRA exhibits substantially lower coverage than the other configurations, particularly for the Gap tier (0.044). Maintenance/when to update is absent from its generated files, while Ownership/contacts appears in only 2\%. These observations suggest that the adapted model does not consistently preserve the requested category structure in this experimental setting.

Overall, the results demonstrate that the benchmark can reveal category-specific generation behavior that would be hidden by aggregate similarity scores.

\paragraph{\textbf{Content Alignment Results}}
\label{sec:content_generation_results}

\begin{table*}[t]
\caption{Content alignment on the AGIT-Instruct evaluation
set. Cos. Sim. indicates Cosine Similarity.}
\label{tab:generation_results}
\centering
\small
\begin{tabular}{lrrrr}
\toprule
& \multicolumn{2}{c}{\textbf{Full-file}}
& \multicolumn{2}{c}{\textbf{Overlap-only}} \\
\cmidrule(lr){2-3}
\cmidrule(lr){4-5}
\textbf{Configuration}
& \textbf{ROUGE 1-F1} & \textbf{Cos. Sim.}
& \textbf{ROUGE 1-F1} & \textbf{Cos. Sim.} \\
\midrule
GPT-4o-mini zero-shot
& 0.423 & 0.681 & 0.386 & 0.710 \\
GPT-4o-mini RAG
& 0.459 & 0.683 & 0.433 & 0.708 \\
CodeLlama-13B Base
& 0.304 & 0.373 & 0.309 & 0.434 \\
CodeLlama-13B QLoRA
& 0.273 & 0.608 & 0.347 & 0.647 \\
\bottomrule
\end{tabular}
\end{table*}
Table~\ref{tab:generation_results} reports the lexical and semantic alignment of generated instructions with developer-written references. GPT-4o-mini RAG achieves the highest full-file ROUGE-1 F1 (0.459), compared with 0.423 for zero-shot generation. Their cosine similarities are nearly identical (0.683 and 0.681). For CodeLlama-13B, QLoRA increases full-file cosine similarity from 0.373 to 0.608, although ROUGE-1 F1 decreases from 0.304 to 0.273. A similar difference is observed for overlap-only evaluation, where QLoRA achieves higher ROUGE-1 F1 (0.347 versus 0.309) and cosine similarity (0.647 versus 0.434).

These results show that different metrics capture different properties of generated content. Notably, the QLoRA configuration achieves higher semantic alignment than CodeLlama Base despite its substantially lower structural coverage. Thus, stronger alignment within generated content does not imply that a model represents the complete template. Conversely, the relatively high structural coverage of the GPT configurations does not establish that all developer requirements are preserved. We therefore complement the automatic evaluation with manual assessment of generated instruction intents.

\paragraph{\textbf{Manual Evaluation Results.}} \label{sec:manual_generation_results}
We manually evaluated 278 RAG-generated category instances against their developer-written references, covering all 16 instruction categories. Among these instances, 54 (19.4\%) fully preserve the reference intent, while 170 (61.2\%) partially preserve it. Together, 224 instances (80.6\%) capture at least some of the intended guidance. However, partial matches substantially outnumber complete matches, indicating that generated instructions frequently capture the general intent without preserving all developer-specified requirements. Another 54 instances (19.4\%) are classified as Missing, while no Wrong instances are observed.

Our analysis reveals that partial matches commonly arise from omitted or generalized requirements. For example, generated project overviews may describe an application's purpose while overlooking its technology stack, documentation, or individual components. Similarly, repository-structure instructions may capture the overall organization but omit specific navigation requirements. Such omissions can make generated instructions appear semantically aligned with their references despite providing incomplete guidance.

These findings highlight an important limitation of automatic content-alignment metrics: lexical or semantic similarity does not necessarily imply complete preservation of developer intent. Although most generated instances capture at least some reference information, only about one in five fully preserves the intended requirements. This suggests that ensuring the completeness of generated instructions remains an important challenge for automated instruction-file generation.

\section{Discussion and Implications} \label{sec:discussion}
Our study extends prior research on agent instruction files by examining their adoption, cross-format consistency, maintenance, and automated generation. We discuss four challenges emerging from these findings and their implications for developers and researchers.

\subsection{Multiple Formats Require Coordinated Maintenance}
Prior studies have characterized the content of agent instruction files and the configuration mechanisms supported by different coding agents
~\citep{Chatlatanagulchai2025AgentREADMEs,
ConfiguringAgenticTools2026,Jiang2026Cursor}.
Our findings extend this understanding by examining how different formats coexist within repositories. Among instruction-bearing repositories in the higher-popularity cohort, 42.2\% maintain multiple formats. Although most same-category sections provide equivalent or complementary guidance (83.9\%), 16.1\% of the 386 examined pairs exhibit divergent guidance, with no directly contradictory pairs identified. Importantly, only 48.2\% of the pairs share the same repository-wide or directory scope. Thus, observed divergence may reflect legitimate component-specific requirements rather than inconsistent instructions. The maintenance challenge is therefore not simply to eliminate differences across formats, but to ensure that instructions applying to the same development context remain coordinated and that complementary guidance is preserved.

\textbf{Challenge and implication.}
The main concern is not simply the existence of multiple files, but keeping their shared guidance consistent as repositories evolve. This concern is reinforced by our maintenance analysis, which finds that 40.4\% of files have only one recorded modification event. Developers should therefore review related instructions across formats when changing build commands, testing procedures, or development conventions. Tool builders could support this process through category-aware comparisons that distinguish equivalent, complementary, and divergent guidance rather than treating textual differences as inconsistencies.

\subsection{Common Categories Do Not Capture All Information Needs}
Chatlatanagulchai et al.~\citep{Chatlatanagulchai2025AgentREADMEs} provided a taxonomy for understanding the information contained in agent instruction files. Ardic et al.~\citep{Ardic2026SCAM} further highlighted variation in the quality and completeness of testing guidance. Our findings complement these studies by showing that category prevalence alone provides an incomplete basis for constructing instruction templates.

In particular, Security/secrets, Ownership/contacts, Maintenance/when to update, and Git/PR conventions have relatively low prevalence but high category addition ratios. This observation motivates our 16-category template, which combines prevalence with category-level maintenance activity.

\textbf{Challenge and implication.}
Developers may overlook less frequently documented information when relying only on existing examples or commonly occurring categories. Our template addresses this challenge by making these categories explicit during authoring and review. However, the Required, Gap, and Recommended tiers represent empirical selection criteria rather than universal importance. Developers should adapt the template to their repository and verify that the included instructions are applicable, actionable, and up to date.

\subsection{Structural Coverage Does Not Guarantee Instruction Quality}
Recent research has examined how repository context and instruction files influence coding-agent behavior ~\citep{Mei2025ContextEngineering,
Lulla2026Impact,Gloaguen2026Evaluating}. Our generation experiments examine a complementary question: whether language models can produce instructions that cover the expected categories and preserve developer-written guidance.

In our exploratory comparison, template-guided generation achieves 82.8\% structural coverage, compared with 18.7\% without guidance. On the AGIT-Instruct evaluation set, GPT-4o-mini zero-shot achieves higher structural coverage than RAG (82.8\% versus 78.6\%), whereas RAG achieves higher full-file ROUGE-1 F1 (0.459 versus 0.423). CodeLlama QLoRA further illustrates this distinction: despite improving embedding similarity over CodeLlama Base (0.608 versus 0.373), its structural coverage is substantially lower (16.6\% versus 64.5\%).

\textbf{Challenge and implication.} Generating the expected categories and preserving their intended content are separate challenges. A model may generate structurally complete instructions with incomplete guidance, or produce semantically similar content while omitting important categories. Therefore, automated instruction generation should not be evaluated using a single aggregate similarity metric. AGIT-Instruct supports complementary evaluation
of category coverage and content alignment, allowing researchers to identify both omissions and weaknesses in generated guidance.

\subsection{Partial Intent Preservation Remains a Generation Challenge}
Our manual assessment provides further evidence of the limitations of automatic generation. Among 278 reviewed RAG-generated category instances, 19.4\% fully preserve the reference intent, 61.2\% partially preserve it, and 19.4\% are missing. These results  cover 16 instruction categories of the proposed template. The predominance of partial matches suggests that generated instructions can capture the general purpose of developer-written guidance while omitting specific requirements. This is particularly important for instruction files, where conditions, commands, and restrictions may determine how an agent performs a task.

\textbf{Challenge and implication.} Fluent and semantically similar instructions are not necessarily complete. Developers should therefore treat automatically generated instruction files as drafts requiring verification, particularly for repository-specific constraints. For researchers, these findings motivate generation approaches that retrieve relevant repository evidence, check individual requirements, and revise incomplete instructions. Future work should also investigate whether improvements in instruction completeness translate into better coding-agent behavior, which our current evaluation does not measure.

Overall, our findings suggest that effective agent instruction files require more than a common structure. Their guidance must remain consistent across formats, evolve with repository changes, and preserve the requirements developers intend coding agents to follow. The proposed template provides an evidence-based authoring structure, while AGIT-Instruct enables systematic evaluation of the remaining generation challenges.

\section{Related Work}\label{sec:related_work}
Research related to our study spans three areas: empirical studies of agent instruction files, the effectiveness of repository-level context for coding agents, and automated generation of software engineering artifacts.

\subsection{Empirical Studies of Agent Instruction Files}
Recent studies have investigated how developers
provide persistent repository guidance to AI
coding agents.
Chatlatanagulchai et al.~\citep{Chatlatanagulchai2025Manifests} examined Claude Code manifests and characterized the information developers provide to agents. Their subsequent study analyzed 2,303 instruction files and developed a taxonomy of 16 content categories~\citep{Chatlatanagulchai2025AgentREADMEs}. They found that developers frequently document functional information, including testing, implementation details, and architecture, while non-functional requirements such as security and performance receive less attention. Other studies have examined individual formats and configuration mechanisms.
Jiang and Nam~\citep{Jiang2026Cursor} investigated Cursor rules and identified recurring themes related to coding conventions, development guidelines, project information, agent directives, and examples. Two follow-up studies confirm and extend these findings: Liang et al.~\citep{RuleTaxonomy2026} present a taxonomy and evolution analysis of rules across AI IDEs, and Sun et al.~\citep{CursorRulesGitHub2026} examine cursorrules files in open-source GitHub projects. Galster et al.~\citep{ConfiguringAgenticTools2026} examined configuration mechanisms across five agentic coding tools, highlighting the widespread use of context files alongside other mechanisms such as skills and subagents.

Recent research has also considered instruction quality and maintenance. Ardic et al.~\citep{Ardic2026SCAM} investigated testing guidance in agent instruction files and identified variation in the completeness of documented testing practices. Santos et al.~\citep{Santos2026ConfigSmells} identified configuration smells in AGENTS.md files, documenting common structural mistakes in how developers configure coding agents. Kapner et al.~\citep{ScanningHarness2026} examined supply-chain defects in AI coding agent configurations, identifying security risks introduced through instruction files. McMillan~\citep{InstructionAdherence2026} investigated how four file-structure variables affect instruction adherence in coding agent configuration files. Voria et al.~\citep{Voria2026Evolution} studied the evolution of agent instructions, contributing to the understanding of these files as maintained software artifacts. The Evolution Context Gap study~\citep{EvolutionContextGap2026} examined missing software evolution information in agent manifests. Related work on agent plans~\citep{AgentPlans2026} investigated task-oriented planning artifacts and the development guidance they provide. Chatlatanagulchai et al.~\citep{PluginMaintenance2026} examined the maintenance and co-evolution of agent plugins in Claude Code plugin marketplaces, providing further evidence that agent configuration artifacts require active maintenance.

These studies establish an important foundation for understanding the content, configuration, and evolution of agent instruction files. Our study extends this research by examining four formats within the same repositories and investigating whether corresponding sections provide consistent guidance. Rather than relying exclusively on content prevalence, we combine category prevalence with category addition ratio (CAR) to derive an evidence-based instruction template. We also examine how this template can support automated instruction generation and evaluation.

\subsection{Effectiveness of Agent Instructions and Repository Context}
A growing body of research investigates the broader landscape of agentic software engineering and whether repository-level instructions improve the performance of AI coding agents. Hassan et al.~\citep{Hassan2025AgenticSE} provide a research roadmap for agentic software engineering, identifying instruction quality as a foundational concern. Watanabe et al.~\citep{Watanabe2025AgenticCoding} studied agentic coding pull requests on GitHub and characterized how AI agents contribute to open-source repositories.
Several studies examine agentic pull requests from complementary
angles~\citep{AgenticPRDevInterv2026, PRMeaningful2026, AgenticPRSecurity2026, AgenticTestsMSR2026, LoggingAgents2026}, documenting patterns in developer interventions, security implications, and code quality outcomes. He et al.~\citep{He2026Cursor} find that Cursor AI increases short-term velocity at the cost of long-term code complexity. Horikawa et al.~\citep{Horikawa2025AgenticRefactoring} study AI coding agents in refactoring tasks. Benchmark suites such as SWE-bench~\citep{Jimenez2024SWEbench} provide standardized evaluations of agent capability on real-world GitHub issues.

Lulla et al.~\citep{Lulla2026Impact} examined the effects of instruction files on agent performance and resource consumption. Gloaguen et al.~\citep{Gloaguen2026Evaluating} evaluated developer-written and automatically generated AGENTS.md files on repository-level coding tasks. Their findings suggest that providing additional instructions does not consistently improve task success and may increase inference cost.

Related research has examined repository context and context engineering for agentic software development~\citep{Mei2025ContextEngineering}. These studies highlight the importance of selecting relevant context rather than simply increasing the amount of information supplied to an agent.

Tafreshipour et al.~\citep{Tafreshipour2025Prompting} characterize how prompts in software repositories evolve over time, showing that developer-written guidance is actively maintained and revised. Studies of documentation pull requests in the agentic era show that documentation artifacts are increasingly co-produced by agents and humans~\citep{DocsAgentHuman2026, LoggingAgents2026}.

Collectively, these findings suggest that instruction-file quality cannot be determined solely by the presence or quantity of guidance. A file may contain relevant categories without providing complete, accurate, or useful instructions. Similarly, generated guidance may appear plausible without preserving the requirements developers intended. Our study examines a complementary aspect of instruction quality. Rather than directly measuring coding-agent task success, we investigate whether automatically generated instruction files represent expected categories and preserve developer-written intents. Through AGIT-Instruct, we evaluate structural coverage, lexical similarity, semantic alignment, and manually assessed intent preservation. These dimensions characterize the generated artifact itself and provide a foundation for future studies connecting instruction quality with downstream agent performance.

\subsection{Automated Generation of Software Engineering Artifacts}
Large language models have increasingly been applied to generating
software artifacts beyond source code, including repository
documentation~\citep{Chakrabarty2025ReadmeReady, Wijaya2025ReadMeLLM},
bug reports and issue-related
artifacts~\citep{Acharya2025BugReports, Choi2025AgentReport,
Nikeghbal2024GIRT}, build
configurations~\citep{Lyu2026DRAFT, Hu2025Repo2Run}, and development
workflows~\citep{Ghaleb2025GitHubActions, ragreflect, shanto2026generating}. These tasks are relevant to agent instruction generation because they require transforming repository information or developer requirements into structured, actionable artifacts.

For repository documentation, Chakrabarty and
Pal~\citep{Chakrabarty2025ReadmeReady} introduced
ReadmeReady, an LLM-based approach for generating
customizable README documentation.
Wijaya et al.~\citep{Wijaya2025ReadMeLLM} proposed ReadMe.LLM to provide LLM-oriented documentation for specialized software libraries. Other research has investigated extracting installation procedures from README files~\citep{readme_installation}, highlighting the challenges of identifying complete development instructions from repository documentation. Empirical studies show that README content and quality affect repository adoption~\citep{READMEPopularity2022}, motivating structured generation approaches. Nikeghbal et al.~\citep{Nikeghbal2024GIRT} introduced GIRT-Model for automated generation of issue report templates, demonstrating that template-guided generation improves structural coverage in software artifacts.

Researchers have also explored the generation and structuring of software maintenance artifacts. Acharya and Ginde~\citep{Acharya2025BugReports} investigated transforming unstructured bug reports into structured reports using LLMs and predefined templates. Choi and Yang~\citep{Choi2025AgentReport} proposed AgentReport, a multi-agent approach for generating reproducible bug reports.

Automated generation has also been investigated for configuration and workflow artifacts. Ghaleb and Rathnayake~\citep{Ghaleb2025GitHubActions} studied the generation of GitHub Actions workflows from natural-language descriptions, identifying challenges involving missing steps and incorrect interpretations. Lyu et al.~\citep{Lyu2026DRAFT} proposed DRAFT, an iterative approach for generating and repairing Dockerfiles using execution feedback. Hu et al.~\citep{Hu2025Repo2Run} introduced Repo2Run for automating repository environment configuration and generating executable Dockerfiles.

These studies illustrate the importance of evaluating generated software artifacts beyond textual fluency. However, agent instruction files differ from executable configuration artifacts: they contain natural-language requirements, conventions, and behavioral constraints that may not have a direct execution-based correctness criterion.

The models and techniques used in AGIT-Instruct build on established work in language model development and fine-tuning. Roziere et al.~\citep{Roziere2023CodeLlama} introduced Code Llama, an open foundation model for code. Parameter-efficient fine-tuning techniques including LoRA~\citep{Hu2022LoRA}, AdaLoRA~\citep{Zhang2023AdaLoRA}, and QLoRA~\citep{Dettmers2023QLoRA} enable adaptation of large models to specialized tasks at reduced cost.

Retrieval-augmented generation~\citep{Lewis2020RAG} provides an alternative approach that grounds generation in retrieved repository content. For embedding-based evaluation, we use Sentence-BERT~\citep{Reimers2019SentenceBERT} to compute cosine similarity between generated and reference instruction sections. Encoder-decoder architectures such as T5~\citep{Raffel2020T5} and CodeT5~\citep{Wang2021CodeT5} represent the class of pretrained models relevant to our generation setup.

Build file maintenance studies~\citep{McIntosh2011Build, McIntosh2015BuildTech} and change complexity analysis~\citep{Hassan2009ChangeEntropy} inform our maintenance metrics and our interpretation of category-level edit patterns across the four
formats.

Unlike prior work focused on generating individual documentation, reporting, or configuration artifacts, our study introduces AGIT-Instruct to evaluate the generation of persistent agent instructions. By combining category-level structural coverage with reference-based content alignment and manual assessment, we investigate whether generated instructions preserve the information developers intend to communicate to coding agents.


\section{Threats to Validity}
\label{sec:threats}

\noindent\textbf{Construct Validity.}
Construct validity concerns whether the
metrics and classifications accurately
represent the concepts being studied.
Classifying instruction content into
predefined categories may introduce
subjective judgments. To mitigate this
threat, two authors independently coded
400 instruction files, achieved a Cohen's
kappa of 0.80, and resolved disagreements
before one author coded the remaining
files. Similarly, differences between
instructions may reflect their intended
scope rather than actual inconsistencies.
We therefore examined the applicability
scope of section pairs and distinguished
equivalent, complementary, and divergent
guidance. For generation evaluation,
structural coverage and textual similarity
may not fully capture instruction
correctness. In particular, converting
instructions into generic intents may
remove repository-specific commands
and constraints. To address this, we
complemented automated metrics with a
manual assessment of intent preservation.
Nevertheless, these measures do not
establish whether generated instructions
are operationally correct.

\noindent\textbf{Internal Validity.}
Internal validity concerns whether
other factors may influence the
observed results. In our template
evaluation, differences in models,
prompts, or repository inputs could
affect the measured improvement.
To mitigate this threat, we used
the same model, base prompt, and
repository inputs in both conditions,
varying only the inclusion of template
guidance. However, variations in LLM
outputs may still influence the results.
In our empirical analysis, differences
between instruction formats may also
reflect repository characteristics
and development practices. We therefore
examined cross-format guidance within
the same repositories and avoided
making causal claims about the
observed differences.

\noindent\textbf{External Validity.}
External validity concerns whether
our findings generalize beyond the
studied repositories, instruction
formats, and models. Our study
examines 3,387 instruction files
from 1,000 GitHub repositories
across four instruction formats
and two popularity cohorts.
However, the findings may not
generalize to private repositories,
other formats, or repositories
outside our dataset. For generation,
we selected 62 evaluation files
to ensure that each of the 16
categories had at least 25 reference
instances. We also separated
repositories across training,
validation, and evaluation to
reduce data overlap. Nevertheless,
this category-based selection may
not reflect the natural distribution
of instruction files, and generation
performance may differ across
other models and repository contexts.

\noindent\textbf{Conclusion Validity.}
Conclusion validity concerns whether
the collected evidence supports
our findings. Our cross-format
consistency analysis relies on a
sample of section pairs rather than
all possible pairs. To reduce
selection bias, we randomly selected
386 pairs from eligible same-category
sections across different instruction
formats within the same repositories,
using a conventional sample-size
target of 95\% confidence and a
5\% margin of error. Two authors
independently evaluated these pairs,
achieving a Cohen's kappa of 0.85.
However, multiple pairs may originate
from the same repository, which may
affect the independence of observations.
Our maintenance analysis also relies
on recorded repository history,
which cannot fully reveal developers'
reasons for modifying instructions.
We therefore interpret modification
frequency, category addition ratios,
and update associations as observed
maintenance patterns rather than
evidence of developer intent or
instruction quality. Finally, although
we evaluate generated instructions
using structural coverage, content
similarity, and manual assessment,
we do not measure their impact on
downstream coding-agent performance.

\section{Conclusion}\label{sec:conclusion}
In this study, we investigated the adoption, content, cross-format consistency, and maintenance of AI agent instruction files by analyzing 3,387 files from 1,000 GitHub repositories across four formats. Our findings show that instruction files are more prevalent in popular repositories and that multiple formats frequently coexist, introducing challenges in maintaining consistent guidance. Among 386 manually examined section pairs, 16.1\% contained divergent information, while 40.4\% of files in our maintenance subset had only one recorded modification event. Based on content prevalence and category addition ratio, we proposed an evidence-based template comprising ten Required, four Gap, and two Recommended categories. We further introduced AGIT-Instruct to evaluate automated instruction generation using structural coverage and content alignment. Our evaluation of four generation configurations on 62 instruction files showed that GPT-4o-mini zero-shot achieved the highest category coverage (82.8\%), whereas RAG achieved the highest full-file ROUGE-1 F1 (0.459). A manual evaluation of 278 generated category instances further reveals that 61.2\% only partially preserve developer intent, highlighting the need to assess instruction completeness beyond structural coverage and automatic similarity metrics. Overall, our findings provide empirical guidance for organizing, maintaining, and generating agent instruction files. Future research should investigate methods for preserving repository-specific requirements, detecting outdated or inconsistent instructions, and evaluating how instruction quality affects downstream coding-agent performance.

\section{Declarations}
\subsection{\textbf{Funding}}
The authors received no specific
funding for this research.
\subsection{\textbf{Ethical Approval}}
Not applicable. 
\subsection{\textbf{Informed Consent}}
Not applicable.
\subsection{\textbf{Author Contributions}}
Mehedi Hasan Shanto: Conceptualization, Methodology, Software, Formal Analysis, Investigation, Visualization, and Writing – Original Draft. Md. Anaytul Islam: Methodology, Data Curation, Investigation, Validation, and Writing – Review \& Editing. Muhammad Asaduzzaman: Conceptualization, Methodology, Supervision, and Writing – Review \& Editing. Alioune Ngom: Supervision and Writing – Review \& Editing.
All authors reviewed and approved the final manuscript.
\subsection{\textbf{Data Availability Statement}}
The datasets, analysis scripts, and replication
materials supporting this study are publicly available at at \url{https://github.com/mehedisresearch/AGIT-Template}.
\subsection{\textbf{Conflict of Interest}}
The authors declare that they have no
conflict of interests.

\bibliographystyle{abbrv}
\bibliography{mybib}

\end{document}